\documentclass[sigconf]{acmart}
\AtBeginDocument{%
  \providecommand\BibTeX{{%
    \normalfont B\kern-0.5em{\scshape i\kern-0.25em b}\kern-0.8em\TeX}}}

\setcopyright{acmcopyright}
\copyrightyear{2018}
\acmYear{2018}
\acmDOI{XXXXXXX.XXXXXXX}

\acmConference[Conference acronym 'XX]{Make sure to enter the correct
  conference title from your rights confirmation emai}{June 03--05,
  2018}{Woodstock, NY}
\acmPrice{15.00}
\acmISBN{978-1-4503-XXXX-X/18/06}
\usepackage{multirow}

\begin{document}

\title{Using Audio Data to Facilitate Depression Risk Assessment in Primary Health Care}

\author{Adam Valen Levinson}
\affiliation{%
  \institution{Yale University} 
 \country{USA}
}

\author{Abhay Goyal}
\affiliation{%
  \institution{Dept. of CS, Missouri S \& T} 
 \country{USA}
}
  
\author{Roger Ho Chun Man}
\affiliation{%
 \institution{Institute for Health Innovation and Technology (iHealthtech), NUS}
 \country{Singapore}
 }

 \author{Roy	Ka-Wei Lee}
\affiliation{%
 \institution{Singapore University of Technology and Design}
 \country{Singapore}
 }

 \author{Koustuv	Saha}
\affiliation{
 \institution{University of Illinois Urbana-Champaign}
 \country{USA}
 }
 
\author{Nimay Parekh}
\affiliation{%
 \institution{Jivox}
  \country{USA}
  }

 \author{Frederick L	Altice}
\affiliation{
 \institution{Yale University}
\country{USA}
 }

\author{Lam	Yin Cheung}
\affiliation{%
 \institution{ORG Research} 
 \country{USA}
 }
 
\author{Munmun De Choudhury}
\affiliation{%
 \institution{Georgia Institute of Technology	}
 \country{USA}
 }

 \author{Navin Kumar}
\affiliation{%
 \institution{ORG Research}
  \country{USA}
}
 \email{navin183@gmail.com}

\renewcommand{\shortauthors}{Kumar et al.}

\begin{abstract}
Telehealth is a valuable tool for primary health care (PHC), where depression is a common condition. PHC is the first point of contact for most people with depression, but about 25\% of diagnoses made by PHC physicians are inaccurate. Many other barriers also hinder depression detection and treatment in PHC. Artificial intelligence (AI) may help reduce depression misdiagnosis in PHC and improve overall diagnosis and treatment outcomes. Telehealth consultations often have video issues, such as poor connectivity or dropped calls. Audio-only telehealth is often more practical for lower-income patients who may lack stable internet connections. Thus, our study focused on using audio data to predict depression risk. The objectives were to: 1) Collect audio data from 24 people (12 with depression and 12 without mental health or major health condition diagnoses); 2) Build a machine learning model to predict depression risk. TPOT, an autoML tool, was used to select the best machine learning algorithm, which was the K-nearest neighbors classifier. The selected model had high performance in classifying depression risk (Precision: 0.98, Recall: 0.93, F1-Score: 0.96). These findings may lead to a range of tools to help screen for and treat depression. By developing tools to detect depression risk, patients can be routed to AI-driven chatbots for initial screenings. Partnerships with a range of stakeholders are crucial to implementing these solutions. Moreover, ethical considerations, especially around data privacy and potential biases in AI models, need to be at the forefront of any AI-driven intervention in mental health care.
\end{abstract}

\ccsdesc[500]{Impact of Computing~Culture}
\ccsdesc[300]{Recognizing and Defining Computational Problems}

\keywords{Audio data, Classification, Depression, Primary Care}


\maketitle

\section{Introduction}
During the COVID-19 pandemic, primary health care (PHC) swiftly transitioned to telehealth. The widespread adoption of telehealth within PHC represented one of the most significant transformations in healthcare brought about by the pandemic. Regulatory adjustments triggered by the pandemic reduced global barriers to telehealth, leading to its widespread expansion. Telehealth is widely regarded as a valuable resource in PHC, with quality of care comparable to or exceeding that of traditional in-person care. 

Major depressive disorder (MDD), commonly known as depression, is a prevalent condition within PHC, affecting up to 20\% of those seeking care \cite{moussavi2007depression}, adding strain to the healthcare system. Depression is a severe mental health condition afflicting over 300 million people worldwide, characterized by feelings of sadness, sleep disturbances, changes in appetite, social withdrawal, and, in severe cases, even suicidal thoughts. Effective treatment of depression may lead to improved patient functioning and reduced healthcare costs. The prolonged untreated duration of depression may negatively impact its course and outcome. PHC serves as the initial entry point into the healthcare system, offering a prime opportunity for depression detection and care initiation. However, even in high-income countries, nearly half of those with recognized depression fail to receive adequate care, representing a significant missed opportunity to address the population-level burden of depression and prevent suicides. Patients generally prefer receiving depression treatment in primary care, and most individuals with depression frequently visit primary care providers. In the context of PHC, the shift to telehealth has notably improved patient attendance for depression diagnosis and treatment, with fewer cancellations and more attended appointments \cite{frank2021mental}. Despite these advancements, over 50\% of potential depression cases remain undetected in high-income countries. Within PHC, about 25\% of diagnoses are either false positives or false negatives, perhaps partly attributed to primary care clinicians prioritizing overall decision-making over mental health diagnoses \cite{aragones2006overdiagnosis}. Furthermore, various barriers may hinder the detection and treatment of depression within PHC, including the misconception that mental disorders are solely the responsibility of specialists, a lack of appropriate tools and guidelines, and low clinician confidence in managing depression, and user-level obstacles. Given the high prevalence and substantial costs associated with depression, focusing on its detection and management within PHC is imperative. Artificial intelligence (AI) may be able to reduce depression misdiagnosis within PHC and improve overall PHC diagnosis and treatment outcomes. 

By harnessing AI, we can enhance the capabilities of telehealth within PHC, equipping clinicians to tackle the intricate challenges of depression diagnosis and treatment. We proposed a proof-of-concept study involving 24 individuals: 12 with MDD diagnoses and 12 without mental health or major health condition diagnoses. The aim was to develop a machine learning model for predicting depression risk using audio data collected from online patient interactions. Previous research successfully employed audio data for depression classification \cite{sardari2022audio}. Telehealth consultations often encounter issues with video, such as poor visual connectivity or dropped calls \cite{wootton2020overcoming}. These issues can negatively affect the relationship between clients and clinicians, reduce client satisfaction, and result in various other challenges. Additionally, audio-only telehealth is more practical for lower-income patients who may lack stable internet connections for audiovisual telehealth \cite{karimi2022national}. Therefore, our study focused exclusively on audio data due to its potential for successful depression classification and the potential absence of video data in some PHC sessions, particularly those involving lower-income patients. Our study had two specific objectives: 1) Gather audio data from 24 individuals; 2) Construct a machine learning model to predict depression risk. The findings from this study could pave the way for the development of machine learning models capable of predicting the risk of various mental health conditions using data from telehealth appointments. 

\section{Related Work}

\textbf{Telehealth in depression treatment}
A meta-analysis indicated that there were no discernible differences in depression severity at post-treatment between those receiving telehealth and those opting for face-to-face care \cite{mitchell2009clinical}. For patients with depression or depressive symptoms, telehealth services may thus serve as a viable alternative to traditional face-to-face care. In another study that examined the shift to mental health-centric telehealth services during the COVID-19 pandemic, researchers explored mental health treatment attendance rates, the delivery of evidence-based interventions, and clinical outcomes \cite{frank2021mental}. Their investigation detailed an improvement in patient attendance following the transition to telehealth. Patients also experienced a decrease in symptoms, and the quality of care remained consistent, with evidence-based interventions consistently utilized over time. Broadly, studies indicated that: 1) Telehealth may be a viable alternative for providing care and treatment to individuals with depression or depressive symptoms; 2) Through telehealth, patients may experience significant improvements and reductions in symptoms around depression. While there were several studies around the use of telehealth in depression treatment, there was limited work on the use of telehealth in depression risk assessment within the PCH setting. 

\textbf{AI in depression diagnosis}
In a recent review article, authors summarized the utilization of AI tools for the automated detection of depression \cite{barua2022artificial}. The authors suggested that AI tools could potentially deliver favorable results and overcome the limitations associated with traditional diagnostic approaches. Furthermore, the review proposed the extraction of a combined set of features, including facial images, speech signals, and visual and clinical history data, for future work aimed at the automated detection of depression and/or anxiety disorders in individuals. Other authors sought to predict depression by employing a novel machine learning pipeline that incorporated biomedical and demographic features from a cohort of 4,184 undergraduate students \cite{nemesure2021predictive}. The resulting model demonstrated an area under the curve (AUC) of 0.67 for depression classification. Notably, the most influential predictive features for depression included satisfaction with living conditions and having public health insurance. These findings highlighted the moderate predictive performance of machine learning methods when applied to the detection of depression using electronic health record (EHR) data. Overall, research in this area detailed that AI tools may be useful for automating the detection of depression. While there seemed to be multiple studies which used AI in the prediction of depression, research did not seem to explore the use of such tools in PCH environments which served mostly marginalized patients. Such environments may favor depression detection that is facilitated by audio-only data. To fill this gap, we conducted a pilot study to determine if audio-only data could predict depression in a telehealth-like setting.  

\textbf{Audio data and depression diagnosis}
Researchers proposed a method for depression detection using audio data, based on a convolutional neural network autoencoder (CNN AE) \cite{sardari2022audio}. The CNN AE extracted relevant and discriminative features from raw audio data, which was then used to classify people as depressed or non-depressed. The authors also used a cluster-based sampling technique to reduce bias towards the major class (non-depressed). Another study developed a model to classify depression using prosodic features (pitch, tone, rhythm) extracted from voice data \cite{negi2018novel}. The model achieved an average F1 score of 0.93. Researchers also developed a technique to conduct depression screening from short voice clips \cite{toto2020audio}. They proposed sliding window sub-clip pooling (SWUP), an audio classification method suited for small datasets. SWUP incorporated elements of deep learning methods with traditional machine learning to achieve higher classification accuracy for smaller datasets while still offering the interpretability of traditional models. SWUP focused on voice sub-clips, thus detecting depression signals as they occurred during speech. The authors achieved an F1 score of 0.74 when predicting depression using only audio data. Studies broadly detailed that: 1) Audio-based depression detection is a promising area of research, as voice data can be easily collected and analyzed using machine learning; 2) Several techniques for audio-based depression detection exist, such as CNN AEs and SWUP. There seemed to be limited studies using solely audio data to facilitate depression diagnosis. To build tools which could assist clinicians in depression diagnosis within PCH, we developed a proof-of-concept study to predict depression diagnosis using audio data. 

\section{Methods}
\textbf{Recruitment}
We recruited potential participants on Facebook groups (e.g., Depression and Anxiety support, Mental Health Awareness and Support), reddit groups (e.g., r/depression, r/Depression Help), and mailing lists (e.g, university student groups) focusing on mental health. A brief overview of the study was provided on these sites, with an email address for participants to contact for further information. Interested participants then contacted study staff via email. We had two distinct sets of inclusion/exclusion criteria as we were trying to recruit two groups of participants: A) Individuals with depression diagnoses; B) Individuals without any mental health diagnoses. For A), those eligible met the following criteria: 1) Greater than 18 years old; 2) Received a depression diagnosis from a clinician, and currently undergoing depression treatment with a clinician; 3) Understood English; 4) Able to provide informed consent. For B), eligible participants had to meet the following criteria: 1) Greater than 18 years old; 2) Have never received any mental health diagnosis, including depression; 3) Currently not receiving treatment for any mental health condition; 4) Understood English; 5) Able to provide informed consent. The following was the exclusion criteria for both groups: Currently seeking or receiving treatment for a major non-mental health condition, such as cancer, heart disease etc. Study staff explained the study protocol to interested individuals and the individual was given a chance to read the consent document and ask questions. If still willing, and if the participant met inclusion/exclusion criteria, full written informed consent was received from the individual, and study procedures would begin. For group A, we asked participants whether they had been diagnosed for depression and were currently undergoing treatment for depression. Participants who answered \textit{yes} to both questions were asked to provide evidence of a formal diagnosis. Examples of such evidence were documents provided by a clinician indicating a depression diagnosis or a insurance document indicating a similar diagnosis. Participants were asked to redact all identifiable information from provided documents. Once depression diagnosis was confirmed, such participants were assigned to group A (depression diagnosis). Those who had never been diagnosed with depression or another mental health condition were assigned to group B (no mental health diagnosis). Given funding constraints, we recruited a total of 24 participants (Diagnosed with depression and currently seeking depression treatment=12; Not diagnosed with depression=12). 

\textbf{Data collection}
Participants were interviewed through a teleconferencing software (Zoom) and their audio was recorded. Zoom is commonly used for telehealth encounters. Participants were first asked a range of demographic questions: Age; Annual income; Gender identity. The interview procedure was designed by two clinicians who worked full-time in PCH, with a diverse caseload. The procedure was developed by the clinicians to be similar to a generic patient encounter. The clinicians independently drafted a set of questions they commonly asked new patients. We compiled all of these questions, 20 in total. We asked all participants these questions in succession. Examples of such questions: Have you had any procedures or major illnesses in the past 12 months? What allergies do you have? Have you served in the military? Are you sexually active? Do you use any kind of tobacco, illicit drugs or alcohol? What prescription medications, over-the-counter medications, vitamins, and supplements do you take? As per past work that assessed depression using voice biomarkers \cite{liu2017speech}, we asked participants to read a short story \cite{mcnamee2004north}. In line with similar past work which developed machine learning techniques to classify depression \cite{shin2021detection}, we collected participant responses to a range of mental health screening tools: PHQ-9, Beck Depression Inventory, Beck Anxiety Inventory, Barratt Impulsiveness Scale. Interviews were about half an hour long. Participants were paid USD25 upon interview completion. The information obtained was recorded in such a manner that the identity of the human subjects could not readily be ascertained, directly or through identifiers linked to the subjects. All collected data was anonymized and encrypted to ensure privacy and security. 
 
\textbf{Data processing}
We adhered to preprocessing steps in previous work \cite{vazquez2020automatic}. We did not need to remove interviewer voices as the teleconferencing software we used provided interviewer and interviewee audio files separately. We removed silences with PyAudioAnalysis \cite{giannakopoulos2015pyaudioanalysis}. Given the small number of audio files and their differing lengths, we cropped all files into shorter segments of four seconds each. 

\textbf{Feature extraction}
Using librosa \cite{mcfee2015librosa}, we extracted the following features from all files: Mel frequency cepstral coefficients (MFCCs); Spectral centroid; Spectral complexity; Zero-crossing rate. MFCC described the power spectrum contour of an audio signal. The audio signal was converted into a frequency domain, using the Discrete Fourier Transform (DFT). Then, the mel-scale was applied to mimic the way humans perceive sound frequencies. From this mel-scaled spectrum, cepstral coefficients were determined. The spectral centroid defined the balance point of the spectrum and is associated with how \textit{bright} a sound appears. Spectral complexity evaluated the variety in the frequency spectrum of the sound signals, helping to differentiate sounds, such as voices, melodies, or background noise. The zero-crossing rate (ZCR) gauged how frequently a signal shifted from positive through zero to negative, or vice versa. MFCCs provided 13 features while the rest provided one feature each, for a total of 16 features per audio file. 

\textbf{Model selection}
We computed all models by randomly splitting (80:20) the dataset into a training and a test dataset. We used the training dataset to find the appropiate machine learning algorithm for our task. To do this, we used TPOT v0.3, an automated machine learning system that selected and preprocessed features based on the training data \cite{olson2016tpot}. The majority of TPOT’s improvements on a range of benchmarks were quite large, with several ranging from 10\%–60\% median accuracy improvement over a random forest analysis. TPOT selected the best machine learning algorithm for our specific task by considering a wide range of feature selection methods and all classification algorithms available in the scikit-learn library. 

\section{Results}
\textbf{Descriptives}
\begin{table}
\centering
\begin{tabular}{|c|c|c|c|c|c|}
\hline
\textbf{Group} & \textbf{Age (M, SD)} & \textbf{Income (M, SD)} \\
\hline
\text{Depression diagnosis} &27.83 (6.86) & 37583.33 (41677.24) \\
\hline
\text{No depression diagnosis} &  29.83 (5.29) & 95416.67 (103781.64) \\
\hline
\end{tabular}
\caption{Participant demographics}
\label{tab:demo_stats}
\end{table}

\begin{table}
\centering
\begin{tabular}{|c|c|c|c|}
\hline
\textbf{Group} & \textbf{Male} & \textbf{Female} & \textbf{Non-Binary} \\
\hline
\text{Depression diagnosis} & 6 & 5 & 1 \\
\hline
\text{No depression diagnosis} & 7 & 5 & 0 \\
\hline
\end{tabular}
\caption{Participant gender identity}
\label{tab:gender}
\end{table}

Average length of the interviews was 26.1 minutes and 25.5 minutes for those with a depression diagnosis, and those without a depression diagnosis, respectively. As indicated, after removing silences, we cropped files into four second segments, resulting in 3969 files (Depression diagnosis=1,632, No depression diagnosis=2,337). We provided participant demographics in Table \ref{tab:demo_stats} and Table \ref{tab:gender}. Age and gender distributions were similar across those with and without a depression diagnosis. Income appeared to be higher for those without a depression diagnosis.

\begin{table}[h!]
\centering
\begin{tabular}{|l|l|l|l|}
\hline
\textbf{Feature} & \textbf{Metric} & \textbf{Depression} & \textbf{No depression} \\
\hline
\multirow{2}{*}{MFCC Mean} & Mean & -74.43 & -75.08 \\
& SD & 7.67 & 7.25 \\
\hline
\multirow{2}{*}{Spectral Centroid} & Mean & 0.0785 & 0.0792 \\
& SD & 0.0261 & 0.0225 \\
\hline
\multirow{2}{*}{Spectral Complexity} & Mean & 37.66 & 29.80 \\
& SD & 31.89 & 31.09 \\
\hline
\multirow{2}{*}{Zero Crossing Rate} & Mean & 0.0541 & 0.0551 \\
& SD & 0.0173 & 0.0199 \\
\hline
\end{tabular}
\caption{Descriptive statistics for features by group}
\label{tab:feature_stats}
\end{table}


In Table \ref{tab:feature_stats}, we presented descriptive statistics regarding features across the participant groups. There was a statistically significant difference in MFCC [t(3967)=2.713, p$<$.05] and spectral complexity [t(3967)=7.752, p$<$.001] between both groups. 


\textbf{Classification}

\begin{table}[h]
\centering
\begin{tabular}{|l|c|}
\hline
\textbf{Metric} & \textbf{Depression} \\
\hline
Precision & 0.98 \\
\hline
Recall & 0.93 \\
\hline
F1-Score & 0.96 \\
\hline
\end{tabular}
\caption{Performance metrics for predicting depression diagnosis}
\label{tab:perf}
\end{table}
Over multiple generations, TPOT evaluated various combinations of preprocessing steps and machine learning algorithms to determine the most effective pipeline based on cross-validation scores. The internal cross-validation scores across generations were as follows: Generation 1: 0.9518; Generation 2: 0.9518; Generation 3: 0.9537; Generation 4: 0.9537; Generation 5: 0.9553. The best pipeline recommended by TPOT achieved an internal cross-validation score of 0.9553. The indicated pipeline involved a K-Nearest Neighbors Classifier with StandardScaler preprocessing as the classifier, with: Number of neighbors: 3, Power parameter for the Minkowski metric: 2. We evaluated and compared the accuracy of the selected model in predicting outcomes using the test data. Table \ref{tab:perf} indicated high performance of the model in classifying depression on all metrics. 

Findings suggested that AI may have the potential to aid in the detection of depression risk using only audio data. We note that this conclusion stems from a preliminary study with a limited sample size. Using machine learning models that rely solely on audio information could be both cost-efficient and simpler to incorporate within Primary Health Care (PHC) settings, especially when compared to models utilizing both audio and visual data. Such audio-focused approaches might be especially valuable in resource-constrained settings, for instance, during phone consultations with economically disadvantaged patients. In areas with unreliable internet access, depending solely on audio data may ensure consistent depression risk assessment. Telehealth platforms used in PHC can be seamlessly integrated with tools such as ours that furnish clinicians with data to assist in diagnosing depression. Incorporating these tools into standard PHC consultations might enhance the accuracy of depression diagnoses, even if the primary reason for the visit is unrelated to mental health. For instance, while discussing a potential flu diagnosis, a clinician might receive a depression risk assessment based on the patient's audio cues, possibly leading to a recommendation for mental health consultation.

\section{Discussion}
\textbf{Implications of Findings} 
Our goal was detail a proof of concept machine learning model for depression risk, using audio data. A strength of our work was the high model performance, keeping in mind limitations of the small dataset. We hope that our work will lead to future work where a larger sample is used to develop tools to predict depression risk on a greater scale. Findings built on past work, demonstrating the use of audio data for depression risk prediction, with potential implications for scaling depression risk classification across marginalized communities. Our findings may lead to a range of possible tools to facilitate depression risk screening and treatment. By developing tools to detect depression risk, patients can be routed to AI-driven chatbots for initial screenings or to provide immediate assistance when human professionals are unavailable. AI-powered chatbots can provide 24/7 support and triage patients. Chatbots can be especially helpful for patients who live in rural or underserved areas, or who have difficulty accessing care during standard business hours. Chatbots can also be used to provide information about mental health conditions, treatment options, and resources. These chatbots can adapted to individual patients, where the chatbots can recognize and respect cultural nuances, ensuring that care is culturally sensitive and relevant \cite{o2023misogynistic}. Broadly, chatbots could make services more accessible without incurring high costs. 

Screening tools may also detect early signs of mental health issues, providing alerts to clinicians when certain patterns are detected. Depression prevalence may be heightened during major events, such as the COVID-19 pandemic. Depression risk detection tools can be used to better understand surges in mental health services demand, helping stakeholders allocate resources more effectively as crises unfold. Similarly, AI can analyze data from PCH telehealth visits to identify at-risk populations and tailor outreach campaigns, or organize AI-driven virtual mental health workshops that are personalized for specific community needs. These workshops can help integrate local mental health practices into standard care, ensuring treatments are culturally rooted and more acceptable. Depression risk identification tools may be also used to determine which patients should receive priority care based on depression severity, allowing clinician schedules to be optimized to maximize the number of patients seen while minimizing burnout. To implement these solutions, partnerships with a range of stakeholders, such as tech companies, local communities, nonprofits, and governments are crucial. Moreover, ethical considerations, especially around data privacy and potential biases in AI models, need to be at the forefront of any AI-driven intervention in mental health care \cite{goyal2023predicting,chen2022categorizing}.

\textbf{Limitations and future work}
Our study involved 24 individuals, which was a relatively small sample size. However, our goal was not to generalize findings, but to demonstrate proof of concept. We recruited participants who had access to the internet and may have not included people without regular internet access. We classified participants into two groups but this approach may not have captured the spectrum of depressive symptoms or other co-existing mental health conditions. While focusing exclusively on audio data may have advantages in low-income populations, we did not obtain video data perhaps useful in diagnosing depression or other mental health conditions. We asked participants to confirm their depression diagnosis through providing documentation, but such a technique may lead to inadvertent sharing of private data. 

Our proof-of-concept study had a relatively small sample size, so future work should collect data from a larger and more diverse group of people. This will make the model more generalizable and robust. We have focused on audio data, but integrating multimodal data (audio, video, text) could provide more insights and improve the model's accuracy. This could also address the limitation of missing out on visual cues. Instead of predicting whether or not someone has depression, future models could predict the severity of depression, from mild to severe. Such a technique would provide a more nuanced understanding of the condition. We also plan to extend our analysis to detect other mental health conditions, such as anxiety and bipolar disorder, based on audio data. Building these products will require the development of a robust ethical and regulatory framework for deploying AI tools. This framework must address concerns about data privacy, consent, and potential biases. We will work with stakeholders in the responsible AI (RAI) space to develop such frameworks. As we will be using private and highly confidential data in future iterations of our tool, we plan to use techniques such as homomorphic encryption to ensure data privacy. Planned work will also involve collaborations with clinicians to refer patients, ensuring verification of a depression diagnosis and minimizing possible loss of personal data. To improve tool development, we will collect feedback from both clinicians and patients. We will also conduct longitudinal studies to understand the usefulness and efficacy of AI tools to predict depression risk. Given rising US healthcare costs, we plan to collaborate with health economists to undertake a cost-benefit analysis of implementing AI in primary healthcare for mental health diagnosis and treatment. This work will aid the rollout of AI systems within large healthcare institutions. Scaling AI systems across health systems will require deep clinical expertise and buy-in from clinicians. Thus, we plan to organize workshops and training sessions for clinicians to familiarize them with our AI tools and ensure they can make the best use of them in their practice. A cornerstone of these workshops would be to seek clinician feedback on tool development and engage clinicians in the core research team \cite{van2023fatherhood,nainani2022categorizing}. By addressing these future research directions, we will be able to better facilitate the use of AI tools in the prediction of depression risk. This could lead to improved detection and management of depression and potentially other mental health conditions within the primary healthcare setting.



\bibliographystyle{ACM-Reference-Format}
\bibliography{sample-base}


\end{document}